# ATOMIC LATTICE DISORDER IN CHARGE DENSITY WAVE PHASES OF EXFOLIATED DICHALCOGENIDES (1T-TaS$_2$)

Robert Hovden[1], Adam W. Tsen[2,3], Pengzi Liu[1], Benjamin H. Savitzky[4], Ismail El Baggari[4], Yu Liu[5], Wenjian Lu[5], Yuping Sun[5,6,7], Philip Kim[8], Abhay N. Pasupathy[2], Lena F. Kourkoutis[1,9]

[1] *School of Applied and Engineering Physics, Cornell University, Ithaca, NY 14853, USA*
[2] *Department of Physics, Columbia University, New York, New York 10027, USA*
[3] *Institute for Quantum Computing and Department of Chemistry, University of Waterloo, Waterloo, Ontario N2L 3G1, Canada*
[4] *Department of Physics, Cornell University, Ithaca, NY 14853, USA*
[5] *Key Laboratory of Materials Physics, Institute of Solid State Physics, Chinese Academy of Sciences, Hefei 230031, People's Republic of China*
[6] *High Magnetic Field Laboratory, Chinese Academy of Sciences, Hefei 230031, People's Republic of China*
[7] *Collaborative Innovation Centre of Advanced Microstructures, Nanjing University, Nanjing 210093, People's Republic of China*
[8] *Department of Physics, Harvard University, Cambridge, Massachusetts 02138, USA*
[9] *Kavli Institute at Cornell for Nanoscale Science, Ithaca, New York 14853, USA*

*Abstract:*
**Charge density waves (CDW) and their concomitant periodic lattice distortions (PLD) govern the electronic properties in many layered transition-metal dichalcogenides. In particular, 1T-TaS$_2$ undergoes a metal-to-insulator phase transition as the PLD becomes commensurate with the crystal lattice. Here we directly image PLDs of the nearly-commensurate (NC) and commensurate (C) phases in thin exfoliated 1T-TaS$_2$ using atomic resolution scanning transmission electron microscopy at room and cryogenic temperature. At low temperatures, we observe commensurate PLD superstructures, suggesting ordering of the CDWs both in- and out-of-plane. In addition, we discover stacking transitions in the atomic lattice that occur via one bond length shifts. Interestingly, the NC PLDs exist inside both the stacking domains and their boundaries. Transitions in stacking order are expected to create fractional shifts in the CDW between layers and may be another route to manipulate electronic phases in layered dichalcogenides.**

*Significance Statement:*
**Low-dimensional materials, like 1T-TaS$_2$, permit unique phases that arise through electronic and structural reshaping known respectively as charge density waves and periodic lattice distortions (PLD). Determining the atomic structure of PLDs is critical toward understanding the origin of these charge ordered phases and their effect on electronic properties. Here we reveal the microscopic nature of PLDs at cryogenic and room temperature in thin flakes of 1T-TaS$_2$ using atomic resolution scanning transmission electron microscopy. Real-space characterization of the local PLD structure across the phase diagram will enable harnessing of emergent properties of thin transition-metal dichalcogenides.**

Layered transition-metal dichalcogenides (TMD), such as $TaS_2$ or $TaSe_2$, are prototypical charge density wave (CDW) systems that spontaneously break lattice symmetry through periodic lattice distortions (PLD). PLDs are associated with dramatic electronic changes such as metal to insulator transitions[1]. Upon cooling from the normal metal phase at > 543 K, the 1T polymorph of $TaS_2$ undergoes several CDW transitions until it finally enters a strongly insulating phase at low temperature where the PLD is commensurate with the crystal lattice[2,3]. In addition to thermal and pressure-induced transitions[4], recent work on thin, exfoliated 1T-$TaS_2$ flakes has demonstrated thickness-tuned conductivity and external electronic control[5]. While CDWs at the surface of bulk crystals have been carefully mapped using scanning tunneling microscopy (STM), less is known about the CDW/PLD structure and stacking order in thin exfoliated TMDs.

Recent theoretical calculations[6-8] and surface measurements[9-11] suggest that the electronic structure of 1T-$TaS_2$ is critically dependent on the CDW stacking order along the c-axis. However, previous work has focused on phase changes of the CDWs alone and variations in atomic lattice stacking were not discussed. Such changes in local topology can have a large influence on the implementation of actual devices based on 2D materials[12]. This became apparent in bilayer graphene, where stacking boundaries dominate the bulk transport behavior[13,14]. The layered transition-metal dichalcogenides have additional complexities—CDW / PLD structure, sensitivity to oxidation[5], and lattice stacking order—that solicit real-space characterization with atomic resolution. Here, we use aberration-corrected and cryogenic scanning transmission electron microscopy (STEM) paired with modern exfoliation techniques to interrogate the PLD structure of thin 1T-$TaS_2$ in both plan-view and cross-section, revealing local variations in PLD coherence across layers and the presence of stacking faults in the atomic lattice. We demonstrate that STEM provides a direct measurement of PLD structures in both room- and low-temperature phases of CDW materials.

A PLD is a spatial modulation of nuclei positions that accompanies the electric field of a CDW[15] and minimizes the crystal's free energy[16,17]. For 1T-$TaS_2$, the atomic sites of the perfect lattice (**r**) are displaced (**r'**) by three modulation waves and harmonics thereof:

$$\boldsymbol{r'} = \boldsymbol{r} + \boldsymbol{A}_{m,n,l} \sin(\boldsymbol{q}_{m,n,l} \cdot \boldsymbol{r} + \phi_{m,n,l}), \qquad \{\boldsymbol{q}_{m,n,l} = m\boldsymbol{q_1} + n\boldsymbol{q_2} + l\boldsymbol{q_3}; m, n, l \in \mathbb{Z}\}$$
(1)

The modulation **q**-vectors are directly visible from reciprocal space peaks via electron diffraction[1,5,18]. To allow direct correlation of the real and reciprocal space structure over smaller domains (~5-100 nm wide) than possible by conventional electron diffraction, here, we use atomic lattice images and their Fourier transforms (FFTs). In addition to the atomic Bragg peaks, the reciprocal space structure exhibits PLD peaks based on their wave vectors, $\boldsymbol{q}_{m,n,l}$, with intensities governed by the amplitude, $A_{m,n,l}$, and phase, $\phi_{m,n,l}$, of each PLD[19,20]. The reciprocal space structure of a single layer is derived in Supplemental Materials.

For 1T-$TaS_2$, the PLD phase transitions correspond to small changes in **q** and **A**[3,21,22]. In the low-temperature commensurate (C) phase (<~183K) the PLD wave vectors are rational fractions of the lattice constant—forming the familiar 'Star of David' with 13 Ta sites (Fig. 1a) in real space. At higher temperatures (280<T<~354K) the wave vectors are nearly-commensurate (NC) with the Bragg lattice–each PLD wave vector ($\mathbf{q}_i$) undergoes a small rotation from ~12° to 13.9° and increases its magnitude by ~2%, although these values are temperature dependent and show hysteresis[3,18].

Here we observe, at atomic-resolution, the PLDs in a transition-metal dichalcogenide—exfoliated 1T-TaS$_2$—with STEM. High-angle annular dark-field (HAADF)-STEM uses elastically scattered high-energy electrons to obtain a projection image of the nuclei's positions in a thin specimen; where intensity scales to first order with the proton number (Z) of each nucleus[23-25]. HAADF-STEM provided atomic-resolution images of the NC and C phases upon in-situ cooling from 293 to 95 K. Experimental details are provided in Supplemental Material.

The transition to the C phase is clearly visible in real space with cryo-STEM, and is marked by the appearance of a low frequency ($\lambda \sim 1.2$nm) commensurate modulation consistent with a $\sqrt{13}\times\sqrt{13}$ supercell (Fig. 1d and Supp. Fig. 3,4). Overlays in Fig. 1d highlight a unit cell in the commensurate PLD. While previous STM results track 1T-TaS$_2$ CDWs at the surface only[26-28], we measure modulations even in projection images of ~65 layer thick regions. Here, the thickness of the flake was determined from convergent beam electron diffraction patterns recorded in the same area (Supp. Material). The visibility of this ~1.2 nm periodicity confirms that the PLD stacking along the c-axis is at least partially ordered for the commensurate phase. For disordered stacking, the supercell would be lost when viewed in projection. Ordering of the CDWs/PLDs in the out-of-plane direction has previously been studied theoretically using free energy calculations[16,29,30]. Compared to the most ordered system, where CDWs are aligned in all layers (Fig. 1b), the energy is lowered by translating the CDW from one Ta site to another Ta site (Fig. 1c) which reduces overlap of interlayer charge[1,16]. Nakanishi & Shiba have predicted several low-energy partially ordered CDW arrangements[30]. For some, the CDW maxima occur only on a subset of the 13 Ta sites (Fig. 1a). In projection such partial ordering would produce periodic superstructures as we observe experimentally.

The PLD is also visible in thinner specimens. Figure 2 shows the same 34 layer TaS$_2$ flake at 293K (Fig. 2a) and 95K (Fig. 2b). As compared to the ~65 layer flake in Fig. 1d, the real-space commensurate PLD modulation ($\lambda \sim 1.2$ nm) in the thinner region appears more disordered (Fig. 2b). The observed increase in PLD disorder in this thinner sample may correlate with previous conductivity measurements that showed that the CDW structure becomes more metastable and the C phase suppressed as the number of TaS$_2$ layers is reduced below ~20 nm[5,31]. Note, our direct imaging results suggest C PLD disorder can exist even in 20 nm flakes. Quantifying the degree of disorder in-plane vs. out-of-plane is, however, hampered by the projection nature of the imaging method.

The transition from the NC to the C phase upon cooling is confirmed in reciprocal space through HAADF-STEM FFTs. Two FFTs of HAADF-STEM images from the same stacking domain in a flake before and after cooling are shown as a composite image in Fig. 2c with the 93K, C-phase in *blue* and the 293K, NC phase in *red*. The six bright peaks (*circled white*) mark the Bragg spots of the atomic lattice. Additional sets of peaks surrounding the central beam correspond to the PLD wave vectors (Supp. Material). Most noticeably, second order harmonic peaks mark the corners of large triangles—three of the large triangles for each phase are drawn over the FFT. The room temperature triangles (*red*) are rotated and smaller than the low temperature triangles (*blue*) confirming the NC-C phase transition. Note, that image compression and distortion artifacts due to sample drift at low temperature hampered perfect alignment of the FFTs in Fig. 2c,d.

The NC-C phase change is more noticeable at low frequencies. In the commensurate phase, a peak occurs at the six first-order **q** positions (Fig. 2d *blue*). The NC phase shows a set of PLD peaks surrounding each first-order **q** position (Fig. 2d *red*) each with reduced intensity compared to the C peaks.

Low-frequency spots in NC 1T-TaS$_2$ have been reported with electron diffraction[1] although often obscured by the zero-beam intensity and have not been discussed. Since the NC PLD wave vectors are not commensurate with the Bragg peaks a vast array of additional spots is geometrically permitted (details in Supp. Mat.)[1,32,33]. We suggest that the array of low-frequency NC peaks observed here (Fig. 2d) constitute high-order PLD harmonics from a first order Bragg reflection. In the C phase, these high-order harmonics cannot be distinguished as they lie at the same position. However, the existence of high-order harmonics in the NC PLD suggests that the first-order peaks of the C phase contain a coherent superposition of high-order harmonics. We observe changes in the appearance of these low-frequency NC satellites between and across thin TaS$_2$ flakes—with asymmetric 'moon' shapes and the appearance of a first order harmonic at the center of three satellites (Supp. Fig. 2d,e,f)—indicating local structure variation. However, tracking structural changes on short length scales such as defects or stacking faults is difficult in Fourier space. These changes are revealed by real-space imaging.

The NC PLD real space structure of 1T-TaS$_2$ was characterized in plan-view and cross-section using aberration-corrected HAADF-STEM. Variations in the planar Ta-Ta interatomic distance were extracted directly from the HAADF image using image analysis based on a peak fitting algorithm (Supp. Mat.). The plan-view NC PLD structure displays visible variations in the Ta-Ta spacing despite being a projection of many layers (Fig. 3a, Supp. Fig 6). This can only occur if the interatomic spacings throughout many layers are correlated. Figure 3a shows local domains in the NC phase (highlighted in green) with boundaries (dark regions) defined by an increased Ta-Ta interatomic spacing—reminiscent of fractured Stars of David. In cross-section (<100> direction) we find the PLD displays local stacking order variations (Fig. 3b). Locally ordered regions containing 3 layer periodicities (Fig. 3d), consistent with the out-of-plane component measured in the FFT (Fig. 3c), were found directly adjacent to disordered regions (Fig. 3e). Thus, the NC PLD is not fully coherent throughout the thickness of the flake and exhibits stacking faults suggesting disorder both in-plane and out-of-plane.

In order to distinguish the PLDs from an amplitude modulation, all Ta centers were determined using 6-parameter Gaussian fits to real space HAADF-STEM data and a Fourier transform was taken of the peak positions only. The presence of superlattice peaks in the resultant FFT can only originate from atomic displacements (Eq. 1) and confirms our assertion of directly measuring PLDs. Using this approach, we observe PLDs in both cross-section and plan-view (Supp. Fig. 6, 7). In the C phase, the atomic displacements due to PLDs were between 11 and 15 pm (Supp. Fig. 3). These displacements are above our measurement precision around 3 pm (Supp. Fig. 4) and consistent with previous bulk x-ray measurement[32], however, higher accuracy may be achievable with recent image acquisition and non-rigid registration schemes[34-36]. The appearance of PLDs remained relatively stable and were not local to regions exposed to the beam; in contrast to other beam-induced restructuring recently observed in related materials (discussion in Supplemental Methods)[37,38].

In addition to disorder in the PLDs, crystallographic stacking domains and their boundaries were revealed in exfoliated 1T-TaS$_2$ without destruction of the CDW. In Fig. 4a, 1T-TaS$_2$ appears as bright Ta atoms in a trigonal array. This is the expected stacking for the perfect 1T polymorph, which we denote as A..A.. (Fig. 4c,g). However, A..B.. stacking domains with a hexagonal array of Ta atoms were also present in the same flake (Fig. 4b). Both stacking arrangements in regions near the stacking boundary contain NC PLD peaks in the FFTs (Fig. 4a,b *inset*).

The boundary between stacking orders is an in-plane translation of one bond length (1.94 Å) that occurs over several nanometers (illustrated in Fig. 4d,h). It is a translation of top and bottom layers separated by a single interlayer fault (hence A..B.. notation) as shown in cross-section (Fig. 4e-h). A stacking boundary spanning ~3 nm is shown experimentally in Fig. 4d. Here, the transition is qualitatively consistent with both compressive and shear strain components (illustrated in Fig 4c) as previously reported for graphene[13]. Geometric phase analysis (Supp. Fig. 11) of this region confirms the transition contains compressive strain ( > 1.5% ) confined to a ~3 nm region. Hexagonal Ta (A..B..) domains in 1T-TaS$_2$ were minority regions but not rare—boundaries were identified in many of the characterized flakes. No fewer than 3 stacking faults were found across the ~7 μm long and less than 50nm thick cross-section sample. In one instance, a hexagonal A..B.. domain transitions to a thin A..A.. domain roughly 15 nm wide, then returns back to hexagonal A..B.. or B..A.. stacking (Supp. Fig. 12,13).

We expect a lattice translation will also translate its CDW structure, creating unique topological properties at the stacking boundary and between the adjacent layers of the stacking fault. The observed A..B.. domains (Fig. 4b) create a fractional translation in the CDW where a layer's CDW peaks (on Ta sites) lie at an adjacent layer's sulfur sites. Previous theoretical[6] and experimental[9,10] results already demonstrated the importance of CDW stacking order on the electronic properties of 1T-TaS$_2$, however, fractional translations were not considered. We expect that the electronic structure will be distinct between the layers at the stacking fault and could dominate the properties in thin films.

In summary, direct imaging of periodic lattice displacements in layered transition metal dichalcogenides (1T-TaS$_2$) revealed the presence of disorder intrinsic to PLDs or crystal stacking. At room temperature we demonstrate that these two forms of disorder can coexist in the system. Compared to surface sensitive techniques such as STM, HAADF-STEM probes the nuclei positions of the entire thin specimen in projection, with particular sensitivity to high-Z elements like Ta.  With cryo-STEM we observed phase transitions in the CDW material—e.g. the NC to C phase transition of exfoliated 1T-TaS$_2$. In the C phase, the PLD superstructure is resolved in projection images of specimens as thick as ~65 layers, suggesting at least partial ordering of the CDW in the out-of-plane direction. Aberration-corrected electron microscopy at room temperature revealed disordered regions in the NC PLD and lattice in both plan-view and cross-section. Furthermore, changes in atomic lattice stacking orders and the corresponding stacking boundaries were resolved at room temperature. These stacking faults should create fractional translations in the CDW and are expected to affect the electronic properties especially in thin films. A systematic study across temperatures, combined with aberration correction, could unveil the kinematics of phase change but will require further instrument optimization stable enough for atomic resolution imaging at a wider range of temperatures. Looking forward, our work opens up the possibility to explore microscopic origins of metastable phases and phase transitions in TMDs using temperature controlled STEM.

ACKNOWLEDGEMENTS

The work at Cornell was supported by the David and Lucile Packard Foundation and made use of the Cornell Center for Materials Research Shared Facilities supported through the NSF MRSEC program (DMR-1120296). The FEI Titan Themis 300 was acquired through NSF-MRI-1429155, with additional support from Cornell University, the Weill Institute and the Kavli Institute at Cornell. Sample fabrication at Columbia University was supported by the NSF MRSEC program through Columbia in the Center for Precision Assembly of Superstratic and Superatomic Solids (DMR-1420634). Salary support was provided by AFOSR (FA9550-11-1-0010, A.N.P.). P.K. acknowledges support from ARO (W911NF-14-1-0638). Y.L., W.J.L, Y.P.S. acknowledges support from the National Key Basic Research and Development Program under Contract No.2016YFA0300404, the National Nature Science Foundation of China (11404342), the Joint Funds of the National Natural Science Foundation of China and the Chinese Academy of Sciences' Large-scale Scientific Facility (U1232139).

AUTHOR CONTRIBUTIONS

R.H., A.W.T, L.F.K. conceived and designed the experiment. Y.L., W.J.L., Y.P.S. synthesized the 1T-TaS$_2$ crystals. A.W.T., I.E. prepared samples. R.H. performed STEM measurements. R.H., P.L., B.H.S. analyzed data. All authors discussed the results and commented on the manuscript. R.H. and L.F.K. wrote the manuscript.

COMPETING FINANCIAL INTERESTS

The authors declare no competing financial interests.

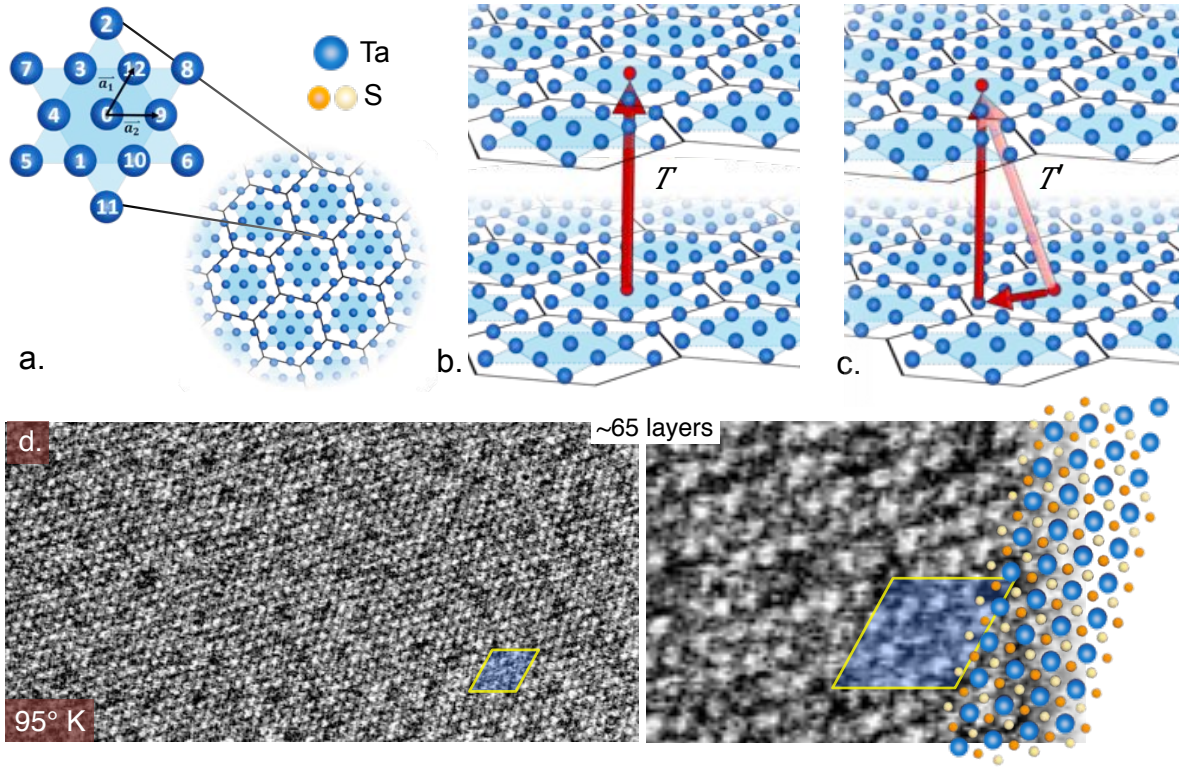

Figure 1 | PLD ordering and atomic resolution HAADF-STEM imaging of thin C phase 1T-$TaS_2$. (a) Within a layer, the commensurate PLD / CDW contains 13 Ta sites in a hexagonal supercell (sulfur not shown). b) Simple stacking of layers results in aligned PLDs with centers (marked red) atop each other. (c) A translation of the PLD between layers can be characterized by stacking vector ***T'*** that connect central Ta-sites (red arrows). (d) At 95K the C-PLD structure is visible in HAADF images of thicker exfoliated flakes suggesting at least partial ordering of the PDLs in the out-of-plane direction. The PLD repeat structure is highlighted by a blue parallelogram with 1.2 nm sides.

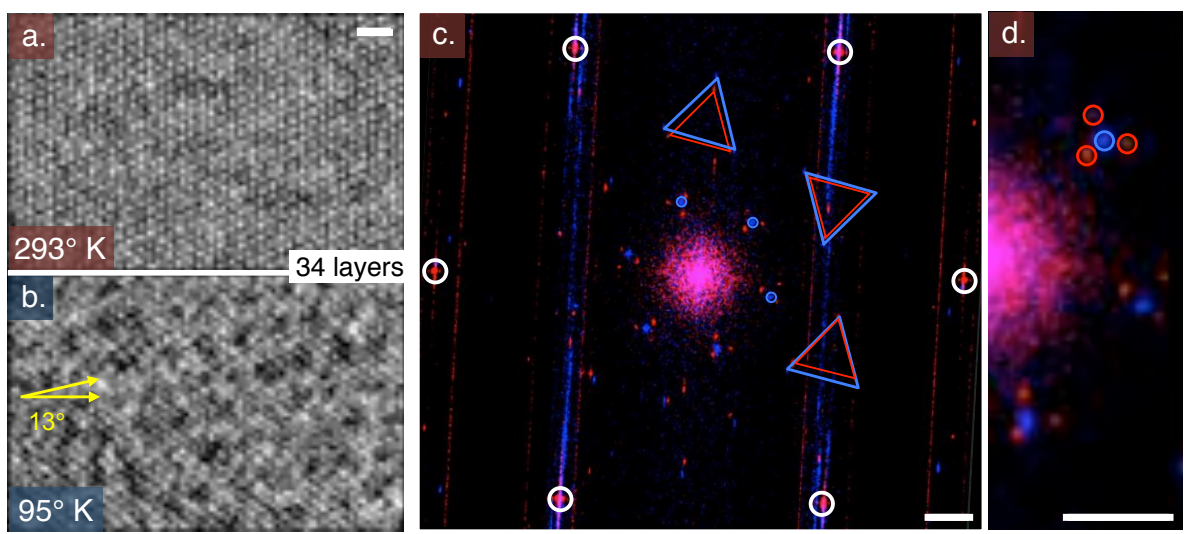

Figure 2 | Atomic resolution HAADF-STEM imaging of NC to C phase transition of thin 1T-TaS$_2$. Upon cooling to 95K, the room temperature phase transitions from the NC (a) to the C phase (b) with C-PLD ordering visible in the HAADF image of 34 layer TaS$_2$. (c) The Fourier transform of NC (red) and C (blue) images show hallmark PLD peaks within the hexagonal Bragg peaks (marked white). (d) The NC phase has sets of three satellites occurring at low frequency, which converge into single spots in the C phase. (c) The second order PLD peaks in the NC phase appear as singular spots that form large triangles (red triangles) that rotate and expand upon commensuration (blue triangles) at lower temperature. Scale bar is 10 Å for images a,b and 0.05 Å$^{-1}$ for c,d. Image processing described in Supplemental Material.

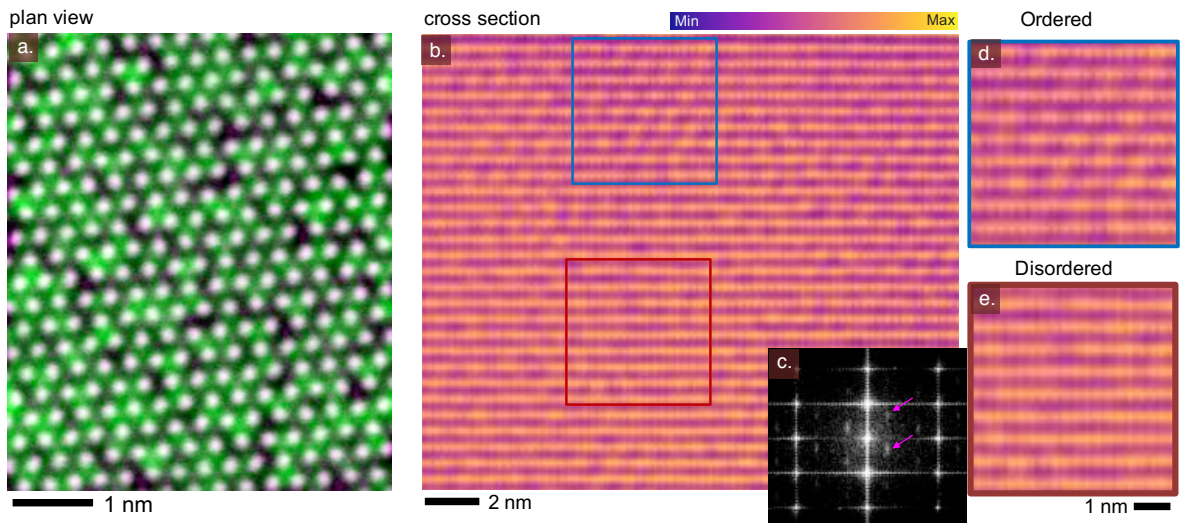

Figure 3 | Disorder in the PLD structure in exfoliated NC 1T-TaS$_2$. (a) In plan-view (<001>), distortions in the atomic positions visualized by superposition of an HAADF-STEM image (grey) with a map reflecting the Ta-Ta interatomic spacing (green). (b) PLD domains in cross-section (<100>) provide visible modulations in real space. (c) Additional peaks in the FFT (0.86 Å$^{-1}$ field of view) show an out-of-plane component to the PLD vector. The strength and coherence of the PLD in cross-section is not uniform, with well ordered (d) and less ordered (e) regions. Cross-section images (b,d,e) have been bandpass filtered with colormap applied to enhance PLD visibility. Image processing described in Supplemental Material.

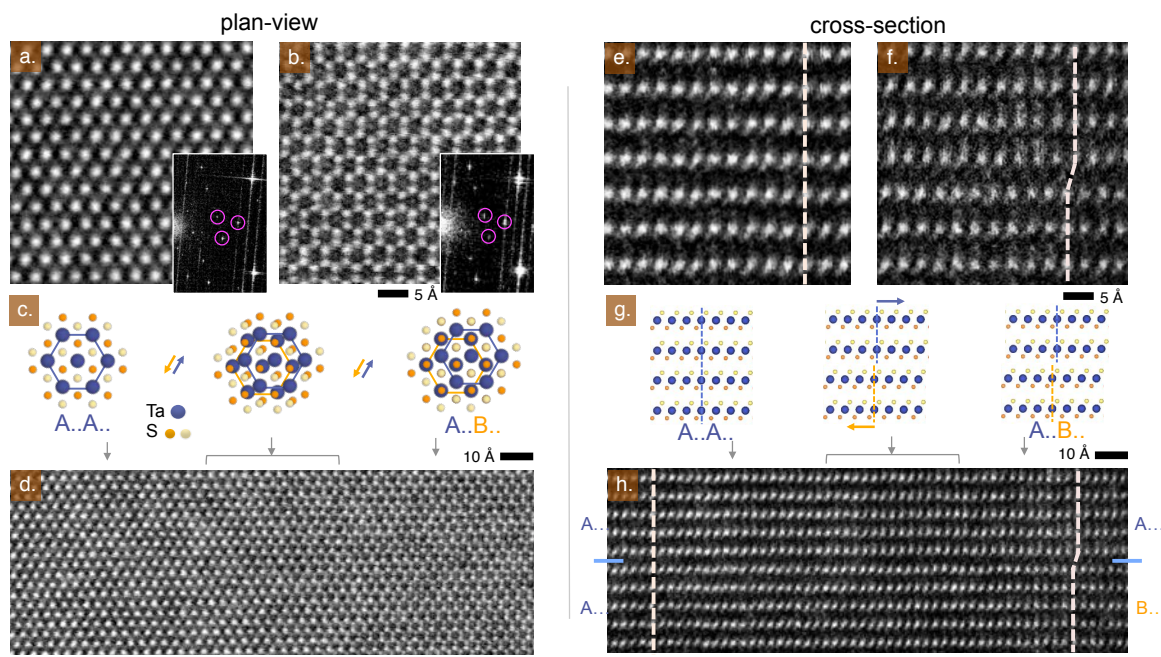

Figure 4 | Atomic imaging of stacking order domain boundaries in thin 1T-TaS$_2$. Bright atoms represent Ta in these HAADF-STEM images. (a,e) Often observed, a trigonal stacking (denoted A..A..) arrangement in which all Ta atoms lie atop each other when viewed along the c-axis (planar vector). (b,f) Hexagonal stacking (denoted A..B..) was also observed with roughly half of the layers having Ta sites shifted one bond length relative to the other layers. Insets (a,b) show a cropped Fourier transform from each specimen region with NC PLD peaks present in both stacking domains. c,g) Illustrations show the stacking arrangements as they transition from A..A.. to A..B.. d) The domain boundary between the two stacking orders transitions continuously over ~30 Å from trigonal (left) to hexagonal (right) stacking. A transition is shown in d) plan-view (<001>) and h) cross-section (<100>).